\documentclass[journal, twoside]{IEEEtran}

\usepackage{cite}

\usepackage{algorithmic}

\usepackage[cmex10]{amsmath}
\usepackage{amssymb,amsthm}
\usepackage{mathrsfs}
\usepackage{bbm}
\usepackage{latexsym}
\usepackage{dsfont}

\usepackage{enumerate}
\usepackage{enumitem}
\usepackage{verbatim}

\usepackage{graphics,graphicx}
\usepackage{color,colortbl}
\usepackage{rotating}

\usepackage{tikz}
\usetikzlibrary{trees}

\newtheorem{theorem}{Theorem}[section]
\newtheorem{lemma}[theorem]{Lemma}
\newtheorem{proposition}[theorem]{Proposition}

\theoremstyle{remark}

\theoremstyle{definition}

\newcommand{\br}{\mathbf{r}}
\newcommand{\bs}{\mathbf{s}}
\newcommand{\bi}{\mathbf{i}}
\newcommand{\bp}{\mathbf{p}}

\newcommand{\Acal}{\mathcal{A}}
\newcommand{\Tcal}{\mathcal{T}}

\newcommand{\Perg}{\mathbb{P}_{\text{\normalfont erg}}}

\usepackage{hyperref}
\graphicspath{{./figures/}}

\newif\ifpreprint
\preprinttrue 	% preprint style
\newif\ifcolor
\colortrue % color images
\begin{document}
\title{Sequential Recurrence-Based Multidimensional Universal Source Coding of Lempel-Ziv Type}

\ifpreprint
\author{Tyll~Krueger, Guido~Mont\'ufar, Ruedi~Seiler, and~Rainer~Siegmund-Schultze% <-this % stops a space
\thanks{T. Krueger is with the Department of Computer Science and Engineering at the Wroc\l aw University of Technology, Wroc\l aw, Poland. E-mail: tyll.krueger@pwr.wroc.pl.}% <-this % stops a space
\thanks{G. Mont\'ufar is with the Max Planck Institute for Mathematics in the Sciences, Leipzig, Germany. E-mail: montufar@mis.mpg.de.}% 
\thanks{R. Seiler is emeritus of the Department of Mathematics at the Technische Universit\"at Berlin, Berlin, Germany. E-mail: ruedi.seiler@integral-learning.de.}% 
\thanks{R. Siegmund-Schultze is with the Department of Mathematics at the Technische Universit\"at Berlin, Berlin, Germany. E-mail: rainer.siegmund-schultze@integral-learning.de.}% 
\thanks{\today.}}
\else
\author{Tyll~Krueger, Guido~Mont\'ufar, Ruedi~Seiler, and~Rainer~Siegmund-Schultze% <-this % stops a space
\thanks{T. Krueger is with the Department of Computer Science and Engineering at the Wroc\l aw University of Technology, Wroc\l aw, Poland. E-mail: tyll.krueger@pwr.wroc.pl.}% <-this % stops a space
\thanks{G. Mont\'ufar is with the Max Planck Institute for Mathematics in the Sciences, Leipzig, Germany. E-mail: montufar@mis.mpg.de.}% 
\thanks{R. Seiler is emeritus of the Department of Mathematics at the Technische Universit\"at Berlin, Berlin, Germany. E-mail: ruedi.seiler@integral-learning.de.}% 
\thanks{R. Siegmund-Schultze is with the Department of Mathematics at the Technische Universit\"at Berlin, Berlin, Germany. E-mail: rainer.siegmund-schultze@integral-learning.de.}% 
\thanks{Manuscript received \today; revised Month Day, Year.}}
\fi

% The paper headers
%\markboth{Journal of \LaTeX\ Class Files,~Vol.~11, No.~4, December~2012}%
\ifpreprint
\markboth{ }%
{Sequential Recurrence-Based Multidimensional Universal Source Coding of Lempel-Ziv Type}
\else
\markboth{Transactions on Information Theory,~Vol.~xx, No.~xx, Month~yyyy}%
{Sequential Recurrence-Based Multidimensional Universal Source Coding of Lempel-Ziv Type}
\fi

\maketitle

\begin{abstract}
We define an algorithm that parses multidimensional arrays sequentially into mainly unrepeated but nested multidimensional sub-arrays of increasing size, and show that the resulting sub-block pointer encoder compresses almost every realization of any finite-alphabet ergodic process on $\mathbb{Z}_{\geq0}^d$ to the entropy, in the limit. 
\end{abstract}


\IEEEpeerreviewmaketitle

\section{Introduction}

\IEEEPARstart{T}{his} paper is about the design of sequential dimension-preserving parsing and coding algorithms for multidimensional arrays of data, with optimality proofs. 

The Lempel-Ziv (LZ) algorithm~\cite{LZ77,LZ78} parses an infinite sequence of symbols sequentially into non-overlapping consecutive blocks, with each block corresponding to the shortest sequence of symbols that has not appeared as a previous block in the parsing. 
The original sequence is then expressed as a sequence of words, each of which is equal to one of the previous words plus an additional symbol. 
Each word is encoded by a pointer to the previously occurring sub-word plus an additional symbol. 
In the limit of infinite sequences, this procedure compresses almost every realization of any stationary process down to its entropy.  
Many lossless data compression algorithms are based on similar recurrence-based parsings and pointers. 

Lempel and Ziv~\cite{2DLZ} showed that the algorithm described above can also be used to compress multidimensional data, 
by first transforming the data to a $1$-dimensional stream. This is done by scanning the data with a space-filling Peano-Hilbert curve that preserves the local correlations. However, this approach results in an encryption of the data correlations, due to the inevitable fractal nature of the scanning curve. 
Furthermore, the local correlations recorded by the scanning curve are essentially enclosed in blocks of side length equal to powers of two. 
In this respect, the algorithm takes correlation-lengths into account that are systematically smaller than optimal. 
Therefore, the convergence rate of this algorithm can be expected to be sub-optimal. 

We propose a resort that leaves the multidimensional structure of the data untouched. 
Our algorithm parses the data into multidimensional blocks. 
As in the LZ algorithm, the blocks are constructed in a way that allows to encode them by pointers to previous blocks plus some additional symbols. 
Here, a main difficulty is to control the number of allowed block shapes together with the amount of block overlaps. 
If multiple block shapes are allowed (which can be useful to avoid overlaps or to maintain the number of additional symbols small), 
then the shape of each block has to be encoded. In the worst case, this leads to a hopeless combinatorial explosion. 
On the other hand, if the blocks are allowed to overlap, then the corresponding portions of the data may need to be encoded multiple times. 

Our algorithm uses cubical blocks of variable size, and controls their overlap by allowing both word repetitions and entirely new words. 
Although in the LZ algorithm all words are different from each other and each new word consists of a previously observed word plus a single additional symbol, 
repeated words or entirely new words are not a fundamental problem for achieving optimal compression. 
All that is needed is that, on the whole, most blocks contain large previously occurring blocks and that their size adjusts to the size of the data array and its entropy. 

Regular parsings with non-overlapping cubical blocks of constant size can be used to obtain optimal codes, whereby  
the size of the blocks has to be chosen depending on the size of the data array and its (empirical) entropy. 
We discuss this approach in Section~\ref{section:blockcode}. 
Such a parsing algorithm is not sequential: 
each time new data entries become available, the block size has to be adjusted to the new data volume and the entire code has to be recomputed. 

In order to overcome this problem, it is desirable to have an algorithm that parses and encodes only the new data entries, 
leaving the old parts of the parsing and encoding unchanged. 
We achieve this by adjusting the block size only for the new data entries. 
We discuss this approach in Section~\ref{section:lzcode}. 
As we will show, this can be done in an asymptotically optimal way by following a simple rule: 
the block size is increased when the proportion of repeated words of the current parsing block size surpasses a pre-specified threshold $0<\phi<1$. 
The dictionaries resulting from this parsing algorithm have cardinalities determined essentially by the entropy of the process generating the data and most of the parsed words contain large nested sub-words from the same dictionaries. 
In this way, our algorithm achieves an asymptotically optimal compression rate for almost every realization of every stationary ergodic process. 

Our approach draws inspiration from various fundamental observations. 
The main source of inspiration is of course the classical LZ algorithm for $1$-dimensional data arrays. 
In addition, we should mention the results by Ornstein and Weis~\cite{OW93}, which relate the recurrence time and the entropy of stationary ergodic processes and show that, in the LZ algorithm, when the string length $n$ is large enough, most of the string is parsed into words of length roughly $\log(n) /h$, where $h$ is the entropy of the process. 
An insightful overview on ergodic theory and recurrence theory in the context of data compression has been given by Shields~\cite{Shields98}. Some of our derivations are guided by his invaluable book on the ergodic theory of discrete sample paths~\cite{Shields96}. 
% \cite{Shields97} % simplistic universal lossy 
Finally, we take advantage of the tools elaborated in our previous work on universally typical sets for multidimensional ergodic sources~\cite{UniversallyTypical}. 

It is worth mentioning that the $1$-dimensional special case of our algorithm is different from the LZ algorithm (our algorithm allows for word repetitions). However, it is possible to regard the LZ algorithm as a limiting case of a modification of our algorithm. 
We provide more details on this relationship in Section~\ref{section:remarks}. 

This paper is organized as follows. 
Section~\ref{section:preliminaries} describes our settings and gives basic definitions. 
In Section~\ref{section:blockcode} we discuss two types of non-sequential universal codes. 
The first one encodes the data by a single pointer to a universally typical library, 
and the second one parses the data regularly into blocks of constant shape and size. 
The main contribution of this work is contained in the following two sections. 
In Section~\ref{section:lzcode} we present our multidimensional sequential recurrence-based algorithm of Lempel Ziv type (Algorithm~\ref{alg3}). In Section~\ref{section:optimality} we prove that this algorithm is universally asymptotically optimal (Theorem~\ref{theorem:optimality}). 
Section~\ref{section:remarks} contains a few final remarks. 

\section{Settings}\label{section:preliminaries}

\newcommand{\Zp}{\mathbb{Z}_{\geq0}}

Consider an alphabet $\mathcal{A}$ of finite cardinality $A=|\mathcal{A}|<\infty$, and consider the non-negative orthant $\Zp^d$ of the $d$-dimensional integer lattice $\mathbb{Z}^d$, for some finite~$d$. 
We denote the set of cubical $n\times \cdots\times n$ arrays with entries from $\mathcal{A}$, called $n$-words, by $\Sigma^n:=\mathcal{A}^{\Lambda_n}$, where 
\begin{equation*}
\Lambda_n :=  \{0,1,\ldots, n-1\}^d \subset \Zp^d.
\end{equation*} 
The set of all cubical arrays is denoted by $\Sigma^\ast := \cup_{n\in\mathbb{N}}\Sigma^n$  
and the set of infinitely extended arrays by $\Sigma := \mathcal{A}^{\Zp^d}$. 

Let $\mathfrak{A}^{\Zp^{d}}$ denote the $\sigma $-algebra of subsets of $\Sigma $ generated by cylinder sets of the form 
\begin{equation*}
[w]:=\left\{ x\in \Sigma :x(\bi)=w(\bi),\bi \in \Lambda \right\}  
\end{equation*}%
for some $w\in \mathcal{A}^{\Lambda }$, $\Lambda\subset\Zp^d$, $|\Lambda|<\infty$. 
Given a subset $C$ of $\mathcal{A}^{\Lambda}$, the corresponding cylinder set is denoted $[C]:=\cup _{w\in C}[w]$. 
Let $\mathbb{P}$ denote the set of probability measures over $(\Sigma,\mathfrak{A}^{\Zp^{d}})$. 

Let $\sigma_{\br}$ denote the natural lattice translation by a vector $\br\in \Zp^{d}$ acting on $\Sigma$ by $\sigma_{\br}x(\bi):=x(\bi+\br)$. 
We use the same notation $\sigma_{\br}$ to denote the induced action on an element $\nu$ of $\mathbb{P}$, $\sigma _{\br}\nu (E):=\nu (\sigma _{\br}^{-1}E)$ for all $E\in\mathfrak{A}^{\Zp^d}$. Here $\sigma_{\br}^{-1}E :=\sigma_{-\br}E$ and $\sigma_{\br}E:=\cup_{x\in E}\sigma_{\br}x$. 
The set of all stationary (translation-invariant) elements of $\mathbb{P}$ is denoted by $\mathbb{P}_{\text{stat}}$, i.e., $\nu \in \mathbb{P}_{\text{stat}}$ if $\sigma_{\br}\nu =\nu$ for each $\br\in \Zp^{d}$. 
The stationary measures $\nu \in \mathbb{P}_{\text{stat}}$ which cannot be decomposed as proper convex combinations $\nu =\lambda _{1}\nu_{1}+\lambda_{2}\nu_{2}$ with $\nu_{1}\neq \nu \neq \nu_{2}$ and $\nu_{1},\nu_{2}\in \mathbb{P}_{\text{stat}}$ are called ergodic. 
The corresponding subset of \ $\mathbb{P}_{\text{stat}}$ is denoted~$\mathbb{P}_{\text{erg}}$. 

We denote by $\nu ^{n}$ the restriction of the measure $\nu$ to the cubical block $\Lambda_{n}$, 
obtained by the projection $\Pi_{n}:x\in \Sigma \rightarrow x^{n}\in \Sigma ^{n}$ with $x^{n}(\bi)=x(\bi)$, $\bi\in \Lambda_{n}$. 
For an arbitrary finite set $\Lambda\subset\Zp^d$, the corresponding projection is defined similarly, $\Pi_\Lambda\colon x\in\Sigma\to x^\Lambda\in\mathcal{A}^\Lambda$.

The entropy rate $h(\nu)$ of a stationary measure \mbox{$\nu\in\mathbb{P}_{\text{stat}}$} is defined as limit of the scaled $n$-word entropies: 
\begin{align*}
H(\nu ^{n}):=& -\sum_{x\in \Sigma ^{n}}\nu ^{n}(w)\log \nu ^{n}(w), \\
h(\nu ):=& \lim_{n\rightarrow \infty }\frac{1}{n^{d}}H(\nu ^{n}).
\end{align*}
Here and in the following we write $\log$ for the base-$A$ logarithm $\log_{A}$. 

A regular $k$-block parsing of $x(\Lambda)\in\Acal^\Lambda$ is the list of sub-arrays $x(\Lambda_k +\br +\bp)$ for all $\br\in k\cdot\Zp^d$ with $\Lambda_k +\br +\bp\subseteq\Lambda$, for some fixed $\bp \in\Zp^d$. 
Forgetting about the relative position of the sub-arrays, 
we identify $\Pi_{\Lambda_{k}+\br+\bp}x\sim \Pi_{k}\sigma_{\br+\bp}x\in \mathcal{A}^{\Lambda_{k}}$. This yields a list of $k$-words. 
We will be mainly interested in parsings where most of $\Lambda$ is covered by blocks $\Lambda_k + \br +\bp$ that are contained in~$\Lambda$. 

The empirical non-overlapping $k$-block probability distribution of an array $x$ over the sites $\Lambda_n$ is defined by the relative frequencies of the distinct $k$-words occurring in the regular $k$-block parsing of $x^{n}$ with $\bp=0$, 
\begin{equation*}
\tilde \mu_x^{k,n}(w):= \frac{1}{\lfloor\frac{n}{k}\rfloor^d} \sum_{\br\in  \Lambda_{\lfloor\frac{n}{k}\rfloor}} \mathds{1}_{[w]}(\sigma_{k \cdot \br} x)\quad \forall w\in\Sigma^k.
\end{equation*} 

% % LONG EQUATION
\newcounter{MYtempeqncnt}
\begin{figure*}[!t]
\normalsize
\setcounter{MYtempeqncnt}{\value{equation}}
\setcounter{equation}{0}
\begin{equation}
C^\mu_m(\delta) : =\{w\in\Sigma^m \colon A^{-m^d(h +\delta)} \leq \mu^m(w) \leq A^{-m^d(h -\delta)}\}. 
\label{equation:entropy-typical-set}
\end{equation}
\setcounter{equation}{\value{MYtempeqncnt}}
\normalsize
\setcounter{MYtempeqncnt}{\value{equation}}
\setcounter{equation}{1}
\begin{equation}
\mathcal{T}^{\mu}_{k}(\delta ,m):=
\Big\{w\in \Sigma ^{k} :  
\!\!\!\!\!
\sum_{\substack{ \br \in m\cdot \mathbb{Z}^{d}:  \\ (\Lambda _{m}+\br+\bp)\subseteq \Lambda _{k}}}
\!\!\!\!\!
 \mathds{1}_{[C^{\mu}_{m}]}(\sigma _{\br+\bp}[w])  
 \geq (1-\delta )\left( \tfrac{k}{m}\right)^{d}\!\! \text{ for some }\bp \in \Lambda _{m}\Big\}.
 \label{eq:longeq1}
\end{equation}
\setcounter{equation}{\value{MYtempeqncnt}}
\hrulefill
\vspace*{4pt}
\end{figure*}

\section{Universal Block Codes}\label{section:blockcode}

In this section we discuss two conceptionally simple but instructive approaches to encode multidimensional data, 
their drawbacks and possible remedies. 
This will serve to prepare key concepts behind our sequential algorithm (Section~\ref{section:lzcode}) and its optimality proof (Section~\ref{section:optimality}). 

\subsection{Universally Typical Sets Coding}

Universally typical sets define universal codes in a natural way. 
Given a universally typical set, 
almost every array can be encoded by its index in that universally typical set. 
Algorithm~\ref{alg1} is an example based on the universally typical sets that we describe in the following. 

% % % % % % % % % % % % % % % % % % % 
\renewcommand{\algorithmicrequire}{\textbf{Input:}}
\renewcommand{\algorithmicensure}{\textbf{Output:}}
\begin{figure}[h] 
\begin{algorithmic}[1]
\REQUIRE data array $x^n$
\ENSURE compressed array 
\STATE  Fix $k=\left\lfloor \sqrt[d]{ \log n^{d}}\right\rfloor$ 
\STATE  Compute the empirical per-site $k$-block entropy \newline$h_x^{k,n} := \frac{1}{k^d} H(\tilde\mu_x^{k,n})$
\STATE Encode $x^n$ by $n$ and its index in $\mathscr{T}_n(h_x^{k,n})$
\end{algorithmic}
\renewcommand{\figurename}{Alg.}
\caption{Universally typical set coding algorithm. }\label{alg1}
\end{figure}
% % % % % % % % % % % % % % % % % % % 

Let $\mathscr{T}_n(h_0)\subseteq\Sigma^n$ denote the set of $n$-words $x^n$ with empirical non-overlapping $k$-block distributions $\tilde\mu_x^{k,n}$ of entropy 
$H(\tilde \mu_x^{k,n})$ 
at most $k^d h_{0}$, with $k= \lfloor \sqrt[d]{\log n^{d}}\rfloor$. % for some $0< c <1$, 
This set is asymptotically universally typical. 
More precisely, 
for each $\mu\in\mathbb{P}_{\text{erg}}$ with $h(\mu) < h_{0}$, the probability $\mu^n(\mathscr{T}_n(h_0))$ tends to one as $n$ tends to infinity. Furthermore, this set has a log-cardinality of order $n^{d}h_{0}$. 
See~\cite[Theorem~3.1]{UniversallyTypical} for a proof of these statements. 
Hence $x^n$ can be encoded by a string of length of order $n^d h_0$. 
We explain the construction more precisely in the following.

Let $h_x^{k,n}:= {H(\tilde \mu_x^{k,n})}/{k^d}$ denote the per-site empirical non-overlapping $k$-block entropy of an $n$-word $x^n$. 
Let the elements of $\mathscr{T}_n(h_0)$ be indexed in order of increasing $h_x^{k,n}$. 
This guarantees that the index of an element $x^n$ in $\mathscr{T}_n(h_0)$ is the same for all $h_0\geq h_x^{k,n}$. 
Now, for any $\mu\in\Perg$, the empirical entropy converges to the true entropy, 
$\lim_{n\to\infty}h_x^{k,n}=h(\mu)$ for $\mu$-almost every $x$. 
See~\cite[Theorem~3.6]{UniversallyTypical} for a proof of this statement. 

In turn, 
for almost every realization $x$ of any $\mu\in\Perg$, 
as $n$ tends to infinity, for any $\epsilon >0$, 
the array $x^n$ can be encoded by its side-length $n$ and its index in 
$\mathscr{T}_n(h_x^{k,n}) \subseteq \mathscr{T}_n(h(\mu)+\epsilon)$. 
The typical set does not need to be included in the code, 
since it can be constructed algorithmically from $n$, $k$, and $h_x^{k,n}$. 
Moreover, we fix $k=\lfloor\sqrt[d]{\log n^d}\rfloor$ and note that $h_x^{k,n}$ can be recovered from $n$ and the index of~$x^n$. 
This results in an optimal code with length of order $n^d h(\mu)$.

A problem of Algorithm~\ref{alg1} is that it requires the entire data array to be fed at once. 
If at a given time, after having compressed the array $x^n$, more data needs to be compressed, 
then the block size $k$, the typical set, and the entire code have to be recomputed. 
Furthermore, although the typical set does not need to be included in the code, constructing it requires exponential running time in~$n^d$, and storing it locally for all $k$, $n$, and $h$ would require enormous resources. These disadvantages disqualify Algorithm~\ref{alg1} for practical purposes.

\subsection{Typical Sampling Sets Coding}

Instead of encoding the entire data array $x^n$ by a single index, a natural approach is to divide $x^n$ into $k$-blocks, and to encode each block by its index in a typical sampling set of $k$-words. 
Demanding that $k$ be at most of log-order $n$, the typical sampling set has an arbitrarily small size, relative to $n$, and can be included in the code. 
Algorithm~\ref{alg2} parses $x^n$ regularly into non-overlapping $k$-blocks, and encodes each block by an index to its first occurrence in the parsing. 
In the following we sketch a proof of the optimality of this algorithm. 

% % % % % % % % % % % % % % % % % % % 
\begin{figure}[h] 
\begin{algorithmic}[1]
\REQUIRE data array $x^n$
\ENSURE compressed array 
\STATE  Fix $k=\left\lfloor \sqrt[d]{\log n^{d}}\right\rfloor$ 
\STATE  Parse $x^n$ regularly into $k$-blocks
\STATE  Encode the list of distinct $k$-words
\FORALL{$k$-blocks}
\STATE Encode word by its index in the $k$-word library
\ENDFOR
\end{algorithmic}
\renewcommand{\figurename}{Alg.}
 \caption{Typical sampling set coding algorithm. }\label{alg2}
 \end{figure}
% % % % % % % % % % % % % % % % % % % 

Given a measure $\mu\in\mathbb{P}_{\text{stat}}$ with entropy rate $h=h(\mu)$ and a positive real number $\delta >0$, 
the level-$\delta$ entropy typical set of $m$-words $C^\mu_m(\delta)$ is defined in~\eqref{equation:entropy-typical-set}. \addtocounter{equation}{1} 

For any $k\geq m$, the $k$-word typical sampling set $\mathcal{T}_k^\mu(\delta,m)$ 
is defined in~\eqref{eq:longeq1}. This is the set of all $k$-words which have a regular $m$-block parsing with at least a $(1-\delta)$-fraction of the resulting words contained in the level-$\delta$ entropy typical set. \addtocounter{equation}{1} % LONGEQUATION1 
As shown in~\cite[Theorem~3.5]{UniversallyTypical} (see Theorem~\ref{thm:sampling0}), this set 
has log-cardinality of order $k^d h(\mu)$. 
Furthermore, for almost every $x$, most $k$-words in the regular $k$-block parsing of $x^n$ belong to the typical sampling set $\mathcal{T}_k^\mu(\delta,m)$. 
More precisely, for appropriately chosen $m$, one has $\tilde{\mu}_x^{k,n}(\mathcal{T}_k^\mu(\delta,m))> 1-\alpha$ for any $\alpha\in(0,1/2)$, when $k$ and $n$ are large enough. 
Hence most of the $k$-blocks of $x^n$ are encoded by pointers to a set of log-cardinality of order $k^d h(\mu)$.  
This implies the asymptotic optimality of Algorithm~\ref{alg2}, in the limit where both $k$ and $n$ tend to infinity, 
with $n$ sufficiently larger than~$k$. 

In Algorithm~\ref{alg2}, one has to include the set of sampled $k$-words in the code. 
In contrast to the universally typical sets used in Algorithm~\ref{alg1}, this set cannot be reconstructed from a few parameters. 
However, choosing $k^d\sim \log n^d$, this set is arbitrarily small, and including it in the code does not affect the compression ratio. 
It is worth mentioning that, if the entropy rate $h(\mu)$ of the process realizing the data is known, 
then a tight upper bound for the block size of an optimal code is $k=\left\lfloor\sqrt[d]{\frac{\log n^d}{h+\epsilon}}\right\rfloor$, $\epsilon>0$. 

Algorithm~\ref{alg2} remedies the problem of dealing with the enormous typical libraries affecting Algorithm~\ref{alg1}. 
On the other hand, it still suffers from the non-sequentiality problem. 
Asymptotic optimality is only guaranteed if the block size $k$ increases with the array size $n$, 
meaning that every time more data needs to be compressed, the entire code has to be recomputed. 

A remedy to this problem is to choose $k$ adaptively, 
depending on $n$, in such a way that only the new data entries are parsed with the updated block size. 
This is precisely the idea of the sequential algorithm that we present in the next section.

\section{Sequential Algorithm}\label{section:lzcode}

\begin{figure}
\begin{center}
\ifcolor
\includegraphics[clip=true, trim=5.6cm 8.9cm 6cm 9.2cm, width=8cm]{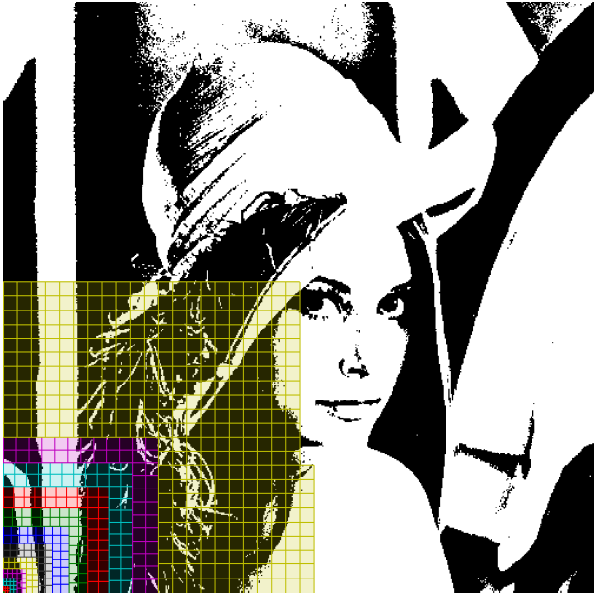}
\else
\includegraphics[clip=true, trim=5.6cm 8.9cm 6cm 9.2cm, width=8cm]{ParsingExampleLenaEps07BW.pdf}
\fi
\end{center}
\caption{
Illustration of a sequential recurrence-based parsing applied to a black-and-white $512\times 512$ pixel image, 
with block size increasing after having parsed a cubical region when the proportion of distinct words with repetitions surpasses a certain threshold. %$\phi=0.07$. 
}\label{figure:example}
\end{figure}
\newcommand{\Bcal}{\mathcal{B}}
\newcommand{\Wcal}{\mathcal{W}}

\newcommand{\SRU}{\operatorname{SRU}}
This section contains the description of our sequential recurrence based parsing and coding algorithm for multidimensional data. 
%, denoted $\SRU_\phi$. 
The algorithm has two components: the first component constructs parsings of multidimensional data based on a word recurrence criterion. 
The second component is a coding scheme based on pointers to sub-words of a given parsing. 
We start with the definition of the coding scheme, in~\ref{section:coding-algorithm}. 
Then we define the parsing scheme, in~\ref{section:parsing-algorithm}. 
An illustration of the parsings generated by our algorithm is given in Figure~\ref{figure:example}. 
In Section~\ref{section:optimality} we will show that the parsing algorithm generates parsings for which the coding algorithm works optimally.

\subsection{The Coding Algorithm}
\label{section:coding-algorithm}

A parsing $P_m$ of an array $x^m = x(\Lambda_m)$ is a list of blocks 
$\lambda_1, \ldots,\lambda_N \subseteq\Lambda_m$ with $\cup_{i=1}^N \lambda_i =\Lambda_m$, 
together with the list of words $w_1=x(\lambda_1), \ldots, w_N = x(\lambda_N)$. 
Clearly, the array $x^m$ is fully described by the words $w_1,\ldots,w_N$ and their relative positions in $\Lambda_m$. 

Given a word $w_l=x(\lambda_l)$, we will consider the set of sub-words of $w_l$, defined by $W_l :=\{ x(\lambda')\colon \lambda'\subseteq \lambda_l \}$. 
Here each $x(\lambda')$ is understood as an element of $\Acal^{\lambda' -\operatorname{min}\lambda'}$, where $\min\lambda'$ denotes the smallest site of $\lambda'$ according to the lexicographic order.\footnotemark %of $\Zp^d$. 
\footnotetext{The lexicographic order of $\mathbb{Z}^d$ is defined by $(\br_1,\ldots, \br_d)<(\bs_1,\ldots, \bs_d)$ iff $\exists j\in\{1,\ldots, d\} \colon$ $\br_i=\bs_i\;\forall i\in\{1,\ldots, j-1\}$ and $\br_j<\bs_j$.}
The sub-words $x(\lambda')\in W_l$ with $\min\lambda'=\min\lambda_l$ are called leading sub-words. % of $w_l=x(\lambda_l)$. 
The smallest site of the block of a leading sub-word is aligned with the smallest site of the block of the containing word.

Having a parsing of $x^m$ at hand, i.e., a list of blocks and words $P_m=(\lambda_1,\ldots, \lambda_N, w_1,\ldots, w_N )$,  
we encode $x^m$ by a sequence $C(x^m) = (C_1,\ldots, C_N)$, 
where $C_i = (p_i,s_i,b_i)$ is the code of $w_i$ and its relative position in $\Lambda_m$. 
The code $C_i$ consists of the following: 
\begin{enumerate}
\item 
A pointer $p_i$, 
which specifies the smallest $j\in\{1,\ldots, i-1\}$ for which 
$w_i(\lambda+\br)=w_j(\lambda+\bs)$ for $\lambda$ having the largest possible cardinality. 
Here $\lambda+\br\subseteq\lambda_i$ and $\lambda+\bs\subseteq\lambda_j$. In case of ambiguity, we choose the lexicographic smallest $\br,\bs\in\Zp^d$. 
If the word $w_i$ does not contain any sub-word of any previous word, we set $p_i=\emptyset$. 
In other words, we consider the largest element in $W_i$ that is contained in some $W_j$ with $j<i$ and set $p_i$ equal to the smallest $j$ for which $W_j$ contains this element. 

\item 
A descriptor $s_i$ of 
$\lambda_i$, $\lambda$, $\br$, $\bs$. That is, a descriptor 
of the shape of $\lambda_i$, its relative position in $\Lambda_m$, 
the shape of $\lambda$, and its relative position in $\lambda_i$ and in $\lambda_{p_i}$. 

\item 
A sequence $b_i$ of the symbols contained in $w_i$ minus the sub-word $w_i(\lambda+\br)$, 
listed according to the lexicographic site order. 
\end{enumerate}

The code $C(x^m)$ is determined uniquely by the parsing $P_m$ and gives a full description of the array $x^m$. 

\begin{figure}
\centering
\ifpreprint
\includegraphics[clip=true,trim=.8cm 1.5cm 1.7cm 0cm, scale=1]{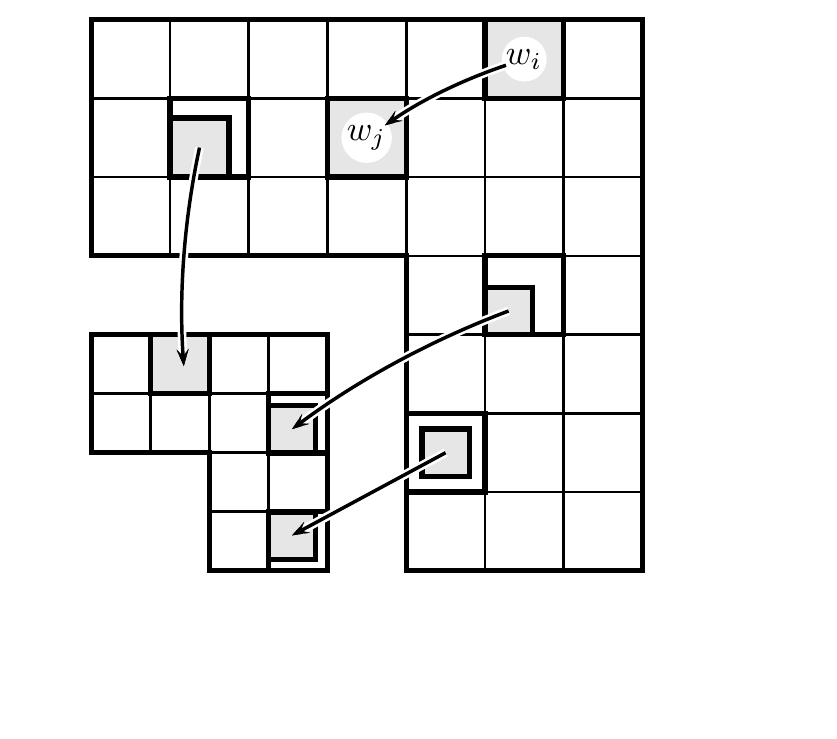}
\else
\includegraphics[clip=true,trim=.8cm 1.5cm 1.7cm 0cm, scale=1]{FigurePointer1}
\fi
\caption{Illustration of the pointers described in~\ref{section:coding-algorithm}. 
Each white square represents a block in a parsing of a two-dimensional array. 
The blocks are enumerated according to the order in which they were parsed, or just according to the lexicographic order of their smallest sites. 
The shaded regions with outbound arrows represent largest sub-blocks of parsed blocks, which hold sub-words of previously parsed words. 
The first occurrences of these sub-words (as sub-words of other parsed words) are shown as shaded regions with corresponding inbound arrows. The last arrow from top to bottom shows an example with non-leading sub-words. 
} \label{fig:pointers}
\end{figure}

In the following we will consider parsings where all blocks $\lambda_i$ are cubical; 
that is, where each $\lambda_i$ is a translate of some $\Lambda_k$. 
Furthermore, we will restrict the set of possible pointers and consider only pointers from leading cubical sub-words to leading cubical sub-words; 
that is, with $\lambda=\Lambda_l$ for some $l\leq k$, and $\br=\bs=0$. 
The pointer structures that we have in mind are illustrated in Figure~\ref{fig:pointers}. 

As already mentioned in the introduction, the quality of the compression algorithm depends crucially on the properties of the parsing. 
A good parsing should consist mainly of words that contain large sub-words of previous words. 
Furthermore, the amount of block overlaps should be as small as possible. 
The parsing algorithm that we describe next produces parsings with the desired properties.

\subsection{The Parsing Algorithm}
\label{section:parsing-algorithm}

Our parsing algorithm has a parameter $0<\phi<1$. 
This parameter will play the role of a threshold for the amount of parsed-word recurrences triggering an increase of the parsing block size. 

Given a parsing $P_m$ of $x^m$, we denote $P_{m,k}$ the list of $k$-blocks in $P_m$, and $P_{m,k}'$ the list of $k$-blocks holding $k$-words that occur at least twice in $P_{m}$. 
We consider the following function that quantifies the amount of $k$-word recurrences in~$P_{m}$: 
\begin{equation}
J(P_{m},k) := |P_{m,k}'|/|P_{m,k}|. \label{condition2} 
\end{equation}
When the parsing $P_m$ does not contain any $k$-words, i.e., $P_{m,k}=\emptyset$, we set $J(P_m,k)=0$. 

\begin{figure}
\centering
\includegraphics[clip=true,trim=.8cm .9cm 1.7cm 0cm, scale=1]{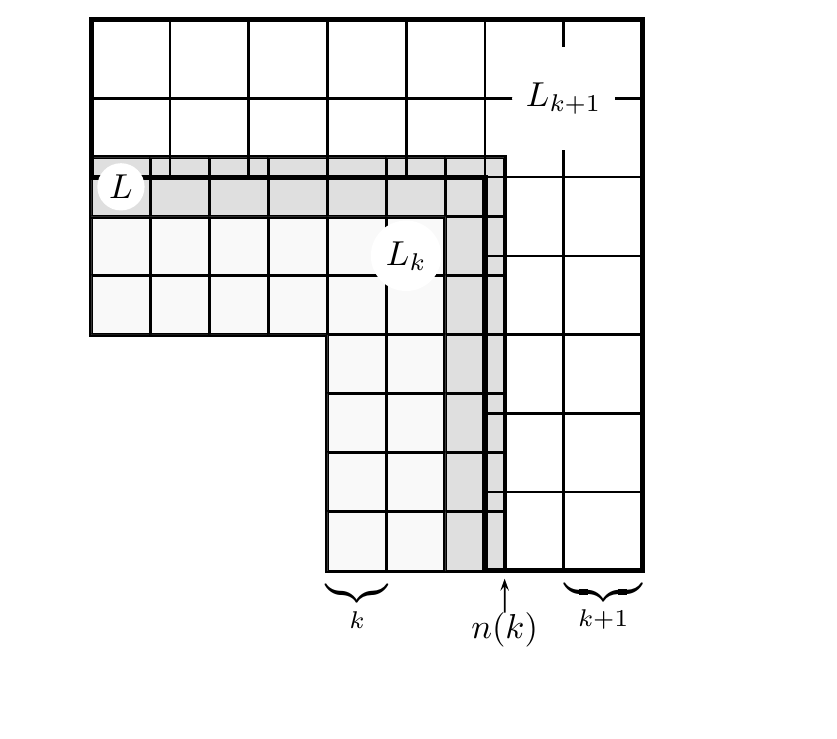}
\caption{Illustration of the parsings described in~\ref{section:parsing-algorithm}. 
The region $L_k$ (shown in gray) represents the set of sites of a two-dimensional array that are parsed with $k$-blocks. 
Within this region, the parsing is regular and has no block overlaps. 
After having parsed the last $L$ boundary region of $L_k$ (shown in dark gray), 
the amount of $k$-word recurrences surpasses the specified threshold $\phi$, i.e., $J(P_{n(k),k})>\phi$, 
and the parsing block size is increased to $k+1$. The first $L$ region of $L_{k+1}$ may overlap with the last $L$ region of $L_{k}$. This happens when $n(k)\operatorname{mod} (k+1)\neq 0$. 
} \label{fig:parsings}
\end{figure}

Our parsing algorithm proceeds as follows: 
\begin{enumerate}
\setcounter{enumi}{-1}
\item 
Fix a recurrence threshold $0<\phi<1$. 

\item 
The input of each iteration is a piece of data $x(\Lambda_n\setminus\Lambda_m)$, $m< n$, 
together with the parsing $P_m$ %or the code $C(x^m)$ 
of $x^m$. 
In the first iteration, $m=0$ and $P_m=\emptyset$. 

\item 
The parsing block size is set equal to the smallest $k\in\mathbb{N}$ for which the amount of $k$-word recurrences in the current parsing $P_m$ is below the threshold $\phi$; that is, the smallest $k$ for which $J(P_m,k) <\phi$. 

\item 
Define the region to be parsed in this iteration. 
For the current $m$ and $k$, this is the $k$-boundary of $\Lambda_m$, defined as 
$L:=\Lambda_{\lfloor\frac{m}{k}\rfloor k +k} \setminus \Lambda_{\lfloor\frac{m}{k}\rfloor k}$. 
This definition allows a small overlap of $L$ and $\Lambda_m$ at the iterations where the parsing block size $k$ has changed. 
%Alternatively, this boundary can be set as $L=\Lambda_{m +k} \setminus \Lambda_{m}$. 
%See Figure~\ref{fig:parsings}. 

\item 
The array region $x(L)$ is parsed regularly into $k$-blocks. 
%When the side-length of $L$ is not divisible by $k$, we allow some overlaps of the $k$-blocks at the `corners' of $L$. 
The blocks in the parsing of $L$ are enumerated according to the lexicographic order of their smallest sites. 

\item 
At this point, the data that still needs to be parsed is $x(\Lambda_n\setminus (\Lambda_m \cup L))$. 
Set $m\leftarrow \left\lfloor m/k\right\rfloor k+k$. 

\item 
Repeat the steps 1--5 until $\Lambda_n$ is exhausted. 
At the last iteration it may happen that $\left\lfloor {m}/{k}\right\rfloor k +k > n$. 
In such a case, set the region $L$ as $\Lambda_n\setminus \Lambda_{n-k}$, 
or, alternatively, set $k=n-m$ and $L=\Lambda_n\setminus\Lambda_m$. 
\end{enumerate}

The parsing structure generated by this algorithm is illustrated in Figure~\ref{fig:parsings}. 
The parsing algorithm, together with the corresponding coding described above, is summarized in Algorithm~\ref{alg3}. 
We denote this compression algorithm $\SRU_\phi$.

% % % % % % % % % % % % % % % % % % % 
\begin{figure}[h] 
\begin{algorithmic}[1]
\REQUIRE data $x(\Lambda_n\setminus\Lambda_m)$ and parsing $P_m$ of $x^m$ with corresponding code $(C_1,\ldots, C_N)$
\ENSURE extension of $P_m$ to a parsing $P_n$ of $x^n$ with corresponding extended code $(C_1,\ldots, C_N, C_{N+1},\ldots, C_{N'})$%
\STATE  $k\leftarrow 1$ 
\WHILE{$m\leq n-k$}
 \IF{$J(P_m,k)<\phi$} 
 \STATE $k$-parse the region $L=\Lambda_{\lfloor \frac{m}{k} \rfloor k + k} \setminus \Lambda_{\lfloor \frac{m}{k} \rfloor k +1}$
 \FOR{each $k$-block $\lambda_i$ in $L$}
 \STATE encode $\lambda_i$ and $w_i=x(\lambda_i)$ by $C_i = (p_i,s_i,b_i)$
\ENDFOR
 \STATE $m \leftarrow \lfloor \frac{m}{k} \rfloor k + k$
 \ELSE 
 \STATE $k \leftarrow k+1$
 \ENDIF
\ENDWHILE
\end{algorithmic}
\renewcommand{\figurename}{Alg.}
 \caption{Sequential recurrence-based multidimensional parsing and coding algorithm $\SRU_\phi$. 
The input declaration illustrates the sequential nature of the algorithm. 
At an initial stage the input may be $x^n$, with $m=0$ and $P_m=\emptyset$.}
\label{alg3}
\end{figure}
% % % % % % % % % % % % % % % % % % % 

\section{Optimality}\label{section:optimality}

Given an array $x^n$ and some fixed $\phi$, 
let $P_n$ denote the parsing of $x^n$ and $C_\phi(x^n)$ the corresponding code of $x^n$ resulting from Algorithm~\ref{alg3}. 
The code $C_\phi(x^n)$ is a string of length $|C_\phi(x^n)|$, with entries from the alphabet $\Acal$. 
Clearly, this $\Acal$-string can be converted to a binary string of length $|C_\phi(x^n)| \lceil\log_2 A\rceil$. 

\begin{theorem}
\label{theorem:optimality}
Let $\mu \in\Perg$ be an ergodic process with entropy rate $h_\mu$. Then, for any $0<\phi<1$,  
\begin{equation*}
\lim_{n\to\infty}\frac{|C_\phi(x^n)|}{ n^d } = h_\mu \quad\text{for $\mu$-almost every $x$}. 
\end{equation*}
\end{theorem}

This theorem shows the universal asymptotic optimality of our compression algorithm (Algorithm~\ref{alg3}).

\subsection{Outline of the Proof}

The proof builds on several propositions and lemmas. 
Lemma~\ref{proposition:typicalsizeL} estimates the typical number of distinct $k$-words that appear in the parsing. 
Proposition~\ref{proposition:lastLs} estimates the relative volume of the portion of $x^n$ that is parsed into words of nearly largest side-length. 
Theorem~\ref{thm:sampling} is a slight generalization of a result from~\cite{UniversallyTypical}. 
It estimates the number of distinct $k$-words that occur in any non-overlapping parsing of a typical realization of an ergodic process. 
Lemma~\ref{lemma:shortpointer} makes a statement about the recurrences of sub-words in the parsings generated by Algorithm~\ref{alg3}.  
It shows that most parsed words contain relatively large leading sub-words which are leading sub-words of previously parsed words. 
With all these tools at hand, we estimate the length of the code, which concludes the proof of Theorem~\ref{theorem:optimality}. 

\subsection{Formal Proof}

In the following, $x$ denotes a realization of an ergodic process $\mu\in\mathbb{P}_{\text{erg}}$ with entropy rate $h_\mu$. 
We consider the parsing of $x$ generated by Algorithm~\ref{alg3} for the recurrence function $J$ defined in~\eqref{condition2} for some fixed threshold $0<\phi<1$. 

We denote by $n(k)$ the side-length of the parsed region right after the block-size is increased from $k$ to $k+1$, 
before the first $(k+1)$-blocks are parsed. See Figure~\ref{fig:parsings}. 

Let $L_k=L_{k}(x)\subseteq \Lambda_{n(k)}$ denote the union of all $L$ boundary regions with parsing block-size $k$. 
%Recall the definition of $L$ given in~\ref{section:parsing-algorithm}~item~3. 

Let $M_k = M_k(x)$ denote the number of distinct $k$-words parsed in $L_k$. 
Let $\tilde M_k=\tilde M_k(x)$ denote the total number of $k$-blocks in the parsing of $L_k$. 

The block-size increase condition $J(P_m,k) \geq\phi$ implies that a $\phi$-fraction of all $k$-blocks in the parsing of $L_k$ hold $k$-words with multiplicity two or more. 
Hence $(1-\phi) \tilde M_k \geq M_k^{(1)}$, where $M_k^{(1)}$ is the number of $k$-words that occur only once in the parsing of $L_k$. 
If $L_k$ is relatively large compared with any of the $L$ boundary regions that it contains, 
then $M^{(1)}_k\approx (1-\phi)\tilde M_k$, and, approximately, $(1-\frac{1}{2}\phi)^{-1} M_k\leq \tilde M_k\leq (1-\phi)^{-1}M_k$. 

The following lemma gives a typical bound for~$M_{k}$. 

\begin{lemma}
\label{proposition:typicalsizeL}
For any $\alpha_0>0$, let 
$B_{\alpha _{0}}^{+}:=\{k:M_{k}\geq A^{k^{d}(h_{\mu }+\alpha _{0})}\}$,  
$B_{\alpha _{0}}^{-}:=\{k:M_{k}\leq A^{k^{d}(h_{\mu }-\alpha _{0})}\}$, 
and $B_{\alpha _{0}}:=B_{\alpha _{0}}^{+}\cup B_{\alpha _{0}}^{-}$. 
Then 
\begin{equation*}
\overline{\mathsf{den}}(\bigcup\limits_{k\in B_{\alpha _{0}}}L_{k}(x))=0\quad\text{for $\mu$-almost every $x$},   
\end{equation*}
where $\overline{\mathsf{den}}(S):=\lim \sup_{n\to\infty} \frac{|S\cap \Lambda_n |}{n^{d}}$ denotes the upper asymptotic density of a set $S\subset \mathbb{Z}_{\geq 0}^{d}$. 
\end{lemma}

\begin{IEEEproof} %[Proof of Lemma~\ref{proposition:typicalsizeL}]
The proof is based on two main ideas. 
First, for each $k\in B^+_{\alpha_0}$, the parsing of $x(L_k)$ contains too many different words, 
contradicting a statement about the number of words needed to construct positive-volumes of realizations of ergodic processes.  %~\cite[Theorem~3.5]{UniversallyTypical}. 
This part of the proof is independent of the criterion $J$ used to construct the parsing. 
Second, for $k\in B^-_{\alpha_0}$, the parsed words repeat so quickly that the array $x(L_k)$ has a very small empirical entropy and thus it allows for an encoding that beats the entropy bound. 
In consequence, these data portions must have a vanishing relative volume. 

{\em Part one. }
Consider first the sets $\bigcup_{k\in B_{\alpha _{0}}^{+}}L_{k}(x)$. 
Assume, contrary to the claim, that there exist a $B>0$ and an integer sequence $(k_i)$ with $k_i\to\infty$ satisfying 

\begin{equation*}
\frac{ \bigcup_{k\in B_{\alpha_{0}}^{+}, k\leq k_i} L_{k}(x) }{|\Lambda_{n(k_i)}|}\geq B \quad  \text{for all $i\in\mathbb{N}$}. 
\end{equation*}

For some $0<B'\leq B$, let $(k_i')$ be a sequence with $k_i'\to\infty$ and $\frac{k_i'}{k_i}\leq 1$, satisfying
\begin{equation*}
\frac{ \bigcup_{k\in B_{\alpha_{0}}^{+}, k_i'\leq k\leq k_i} L_{k}(x) }{|\Lambda_{n(k_i)}|}\geq B' \quad  \text{for all $i\in\mathbb{N}$}. 
\end{equation*}

Let $(k_i^\ast)$ be a sequence with  $k_i^\ast\to\infty$ and $\frac{k_i^\ast}{k_i'}\to 0$. 
Consider a regular $k_i^\ast$-block parsing of $\bigcup_{k\in B_{\alpha_{0}}^{+}, k_i'\leq k\leq k_i} L_{k}(x)$. 
For each $k\in B^+_{\alpha_0}$, $k_i'\leq k\leq k_i$, the region $L_k$ has at least $M_k\geq A^{k^d(h+\alpha_0)}$ different $k$-words, and hence the number $E$ of different $k_i^\ast$-words in the parsing satisfies 
\begin{equation*}
E^{\left\lfloor \frac{k}{k_i^\ast} \right\rfloor^d} A^{k_i^\ast 2d k^{d-1}} \geq M_k \geq A^{k^d(h+\alpha_0)}. 
\end{equation*}
Here, the first term in the left hand side is the number of cubical side-length $\left\lfloor {k}/{k_i^\ast} \right\rfloor^d$ arrays with entries given by $k_i^\ast$-words from a set of cardinality $E$. The second term in the left hand side is an upper bound on the number of ways in which the region $\Lambda_k \setminus \Lambda_{\left\lfloor {k}/{k_i^\ast} \right\rfloor k_i^\ast}$ can be filled with entries from $\mathcal{A}$. 
In turn,  
\begin{equation*}
\log E 
%\geq \left(  k^d(h + \alpha_0) - k_i^\ast 2d k^{d-1} \right)\! \left\lceil \tfrac{k}{k_i^\ast} \right\rceil^{-d} 
\geq (k_i^\ast)^d (h +\alpha_0/2). 
\end{equation*}
This shows that a regular $k_i^\ast$-block parsing of a $B'$-portion of $\Lambda_{n(k_i)}$ contains at least $A^{(k_i^\ast)^d (h+\alpha_0 / 2)}$ different $k_i^\ast$-words.  

On the other hand, by~\cite[Theorem~3.5]{UniversallyTypical} (see Theorem~\ref{thm:sampling0} below), for any $\alpha>0$, there is a set $\mathcal{T}_k = \mathcal{T}_k(\alpha)$ with $ \log|\mathcal{T}_k| \leq {k^d}(h +\alpha)$, which suffices to uniformly construct at least a $(1-\alpha)$-portion of $x^n$, when $k$ and $n/k$ are large enough, for $\mu$-almost every $x$. 
Choosing $\alpha< \frac{\alpha_0}{2}$, $\alpha< B'$, as well as $k_i^\ast$ large enough, this is a contradiction. 

{\em Part two. }
Assume now that there is a $B>0$ and a sequence $(k_i)_i$ with $k_i\to\infty$, satisfying 
\begin{equation*}
\frac{ \bigcup_{k\in B_{\alpha_{0}}^{-}, k\leq k_i} L_{k}(x) }{|\Lambda_{n(k_i)}|}\geq B \quad  \text{for all $i\in\mathbb{N}$}. 
\end{equation*}
For each $k$, let $G_k=G_k(x)$ denote the union of all $k$-blocks in the parsing of $L_k(x)$ that hold $k$-words occurring for the first time in the parsing. Furthermore, let $\bar{G}_k = L_k\setminus G_k$. 
By the construction of the parsing, $|\bar{G}_k|\geq \frac{\phi}{2} |G_k|$. 
Choose a sequence $(l_i)_i$ with $l_i\to\infty$ and $\frac{l_i}{k_i}\to 0$. 
Let $\bar G^i = \bigcup_{k\in B_{\alpha_{0}}^{-}, k\leq k_i} \bar G_{k}(x)$. 
Let $\gamma_i:=|\bar G^i|/|\Lambda_{n(k_i)}|\geq B\frac{\phi}{2}$. 
We can encode the region $G^i = \Lambda_{n(k_i)}\setminus \bar G^i$ of the array $x$ with an optimal $l_i$-block coding algorithm, 
which results in a code of length at most $(1-\gamma_i) (n(k_i))^d (h+\tilde\alpha)$, 
whereby $\tilde \alpha$ can be chosen arbitrarily close to zero for $n(k_i)$ large enough. 

The region $\bar G^i$ consists of repetitions of words that occur in the region $G^i$. 
Hence, by using pointers to the words in $G^i$, the region $\bar G^i$ can be encoded by a string of length at most 
$\sum_{k\in B_{\alpha_0}^-, k\leq k_i} \log(M_k) \frac{\bar{G}_k}{k^d}$. 
Here $\bar{G}_k / k^d$ is the number of pointers, $\log(M_k)$ is the length of each pointer, and we omitted lower order terms. 
By the assumption on the cardinality of $M_k$ for $k\in B_{\alpha_0}^-$, 
we obtain the upper bound $\sum_{k\in B_{\alpha_0}^-, k\leq k_i} k^d (h -\alpha_0) \bar{G}_k / k^d  = \sum_{k\in B_{\alpha_0}^-, k\leq k_i} (h-\alpha_0) \bar{G}_k$. 

In total, we can produce a code of $x(\Lambda_{n(k_i)})$ of length $(1-\gamma_i)(n(k_i))^d(h+\tilde\alpha) + \gamma_i (n(k_i))^d(h-\alpha_0)\leq  (n(k_i))^d(h-\alpha_0')$, for some $\alpha_0'>0$. 
This is a contradiction to the non-existence of codes beating the entropy bound infinitely often. 
\end{IEEEproof}

The next proposition shows that, asymptotically, 
most of $\Lambda_n$ is parsed in words of size close to the largest $k$ that is typical in the sense of Lemma~\ref{proposition:typicalsizeL}. 

Consider the sequence $K= (k^\ast_i)$ consisting of all $k\in\mathbb{N}$ for which $L_k$ has the typical number of distinct parsed words;  
that is, $A^{h_\mu k (1-\alpha_0)}\leq M_k \leq A^{h_\mu k (1+\alpha_0)}$ for some $\alpha_0>0$. 

For any two integers $l\leq u$, let $|L_l^u| = \sum_{l \leq k \leq u}|L_k|$ denote the volume of all $L_k$ regions that are parsed with blocks of side-length between $l$ and $u$, counting overlaps. 

\begin{proposition}
\label{proposition:lastLs}
For any fixed $\alpha >0$, the quotient $|L_1^{(1-\alpha)k_i^\ast}| / |L_{(1-\alpha)k_i^\ast}^{k_i^\ast}|$ tends to zero as $i$ tends to infinity.   
\end{proposition}

\begin{IEEEproof} %[Proof of Proposition~\ref{proposition:lastLs}]
Fix some $\alpha_0 < \alpha$. 
By Lemma~\ref{proposition:typicalsizeL}, for all~$i$, 
 \begin{align}
\sum_{k \leq (1-\alpha) k^\ast_i } |L_k| 
&=\sum_{\begin{smallmatrix} k\not\in K\colon \\ k \leq (1-\alpha)k^\ast_i\end{smallmatrix}} \!\!\! |L_k| 
+ \sum_{\begin{smallmatrix}k\in K\colon \\ k \leq (1-\alpha)k^\ast_i\end{smallmatrix}} \!\!\! |L_k|\nonumber\\
&\leq (1+ o(1)) \!\!\!\sum_{\begin{smallmatrix} k\in K\colon \\ k \leq (1-\alpha)k^\ast_i\end{smallmatrix}} \!\!\! A^{k^d(h_\mu+\alpha_0)} \nonumber\\
&\leq (1+ o(1)) C A^{(1-\alpha)^d(k^\ast_i)^d(h_\mu+\alpha_0) +1} \nonumber \\
&\leq  C' A^{(1-\alpha)^d(k^\ast_i)^d(h_\mu+ 2\alpha_0)}, \label{eq:upperbd}
\end{align} 
for some constants $C$ and $C'$. 
On the other hand, 
\begin{equation}
\sum_{(1-\alpha)k^\ast_i  \leq k \leq  k^\ast_i } |L_k| \geq A^{(k^\ast_i)^d(h_\mu -\alpha_0)}.
\label{eq:secondsum}
\end{equation} 
Choosing $\alpha_0$ sufficiently small, e.g., satisfying $\alpha_0<\alpha h_\mu /3$, we have that $(h_\mu-\alpha_0)>(h_\mu+2\alpha_0)(1-\alpha)^d$, and  the right hand side of~\eqref{eq:secondsum} is exponentially larger than the right hand side of~\eqref{eq:upperbd}. 
\end{IEEEproof}

Before proceeding, we recall the following. 
\begin{theorem}[Theorem~3.5a in~\cite{UniversallyTypical}]
\label{thm:sampling0}
Let $\mu\in\Perg$ with entropy rate $h(\mu)=h$ and let $\delta\in(0,\frac{1}{2})$. Then, for all $k$ larger than some $k_0=k_0(\delta)$ there is an $n_0=n_0(k,\delta)$ such that, if $n\geq n_0$, 
there is a set $\mathcal{T}_k\subseteq\Sigma^k$ of log-cardinality
\begin{equation*}
\frac{\log|\mathcal{T}_k|}{k^d} \leq h + \delta, 
\end{equation*}
with 
$\tilde \mu_x^{k,n}(\mathcal{T}_k) > 1-\delta$ for $\mu$-almost every $x$. 
An example of $\Tcal_k$ are the typical sampling sets $\Tcal^\mu_k(\delta/2 , m)$ from~\eqref{eq:longeq1} with $m\xrightarrow{k\to\infty}\infty$ and $m=o(k)$. 
\end{theorem}

The next theorem is a slight generalization of Theorem~\ref{thm:sampling0}. 
It states that, asymptotically, most words in any non-overlapping $k$-block parsing of any positive portion of $x^n$ (instead of a regular parsing of all $x^n$) belong to a set of log-cardinality close to $h(\mu) k^d$, almost surely.

\begin{theorem}\label{thm:sampling}
Let $\mu \in \Perg$ with entropy rate $h(\mu)=h$ and let $\delta>0$, $\delta'>0$. 
Consider a collection of non-overlapping $k$-blocks $\lambda_1,\ldots, \lambda_I$ covering a region $\Xi\subseteq\Lambda_n$ with $|\Xi|/|\Lambda_n|\geq \beta >0$. Then, for all $k$ larger than some $k_0=k_0(\delta,\delta')$ there is an $n_0=n_0(k,\delta,\delta')$ such that, if $n\geq n_0$, 
there is a set $\mathcal{T}_k\subseteq\Sigma^k$ of log-cardinality  
\begin{equation*}
\frac{\log |\mathcal{T}_k|}{k^d} \leq h + \delta, 
\end{equation*} 
with $w_i=x(\lambda_i)\in \mathcal{T}_k$ for at least a $(1-\delta')$-fraction of all $i\in I$, for $\mu$-almost every $x$. 
Here, the set $\mathcal{T}_k$ does not depend on the specific $x$. 
\end{theorem}

\begin{IEEEproof} %[Proof of Theorem~\ref{thm:sampling}]
The proof follows the steps of the proof of {\cite[Theorem~3.5]{UniversallyTypical}}. 
We show that $\Tcal_k$ can be chosen as the typical sampling sets $\Tcal^\mu_k(\delta/2 , m)$ from~\eqref{eq:longeq1} with $m\xrightarrow{k\to\infty}\infty$ and $m=o(k)$. 

Fix some $\delta$, ${\delta'}$, and $\beta$. 
The Shannon-McMillan-Breiman theorem for amenable groups by Ornstein and Weiss~\cite{OrWe83} states that\footnote{Here in fact we only need the convergence in probability, shown in~\cite{Kieffer75}, ensuring $\mu(C_m^\mu)\xrightarrow{m\to\infty}{1}$.}
\begin{equation*}
\lim_{m\to\infty} -\frac{1}{m^d} \log \mu^m(x^m) = h(\mu),\quad \text{$\mu$-almost surely. }
\end{equation*}
Thus, by the definition~\eqref{equation:entropy-typical-set} of the entropy typical sets $C^{\mu}_{m}(\delta)$, 
there exists an $m_{0}=m_0(\delta,\delta', \beta, x)$ 
such that $\mu^{m}\left( C^{\mu}_{m}\right) \geq 1- {\delta'} \delta \beta /8$ for all $m\geq m_{0}$. 
Fix such an $m\geq m_0$. 
The individual ergodic theorem~\cite{Lindenstrauss:2001} guarantees that the following limit exists for $\mu$-almost every $x$: 
$$\lim_{n\rightarrow \infty }\frac{1}{n^{d}}\sum_{r\in \Lambda _{n}}\mathds{1}_{[C^{\mu}_{m}]}\left( \sigma _{r}x\right) =
\int \mathds{1}_{[C^{\mu}_{m}]}(x)d\mu(x)=\mu^{m}(C^{\mu}_{m}).$$  %
Therefore, the inequality 
\begin{align}
\sum_{r\in \Lambda _{n-m+1}}\mathds{1}_{[C^{\mu}_{m}]}(\sigma _{r}x)
 &\geq (1-{\delta'}\delta\beta/7)(n-m+1)^{d} \nonumber\\
 &> (1-{\delta'}\delta \beta/6)n^{d}  \label{firproperty}
\end{align} 
holds eventually almost surely, i.e., for $\mu $-almost every $x$ and $n\geq n_0(x,m)$. 

Consider an $x$ and an $n\in \mathbb{N}$ for which~\eqref{firproperty} is satisfied. 
Choose a $k$ with $m<k<n$, and consider the non-overlapping $k$-blocks $\lambda_1,\ldots, \lambda_I$ that cover an $\beta$-fraction of $\Lambda_n$. 
If $k/m$ and $n/k$ are large enough, then 
at least $(1-{\delta'} )I$ of the words $w_1,\ldots, w_I$ satisfy 
\begin{equation}
\frac{1}{(k-m+1)^{d}}\sum_{\mathbf{s}\in {\Lambda}_{k-m+1}}\mathds{1}_{[C^{\mu}_{m}]}(\sigma_{\bs}[w_i])
 \geq  (1-\delta /4), 
\label{property}
\end{equation}%
where $[w_i]:=\{x\in\Sigma \colon x(\bi)=w_i(\bi), \bi\in\Lambda_k \}$. 

To see this, note that if more than ${\delta'} I$ of the $w_i$ had more than a $\delta /4$-fraction of `bad' $m$-blocks (blocks with contents not in $C_m^\mu$), then the total number of `bad' $m$-blocks in $x^n$ would be larger than 
${\delta'} I \frac{\delta }{4} (k-m+1)^{d}$. 
For $n/k$ and $k/m$ large enough, this is lower bounded by 
${\delta'} \beta \left\lfloor\frac{n}{k}\right\rfloor^d\frac{\delta }{4} (k-m+1)^{d}
\geq {\delta'} \beta \frac{\delta}{6}n^d$, 
%\end{equation*}
contradicting~\eqref{firproperty}. 

Now, if $k\geq 8 dm/\delta$, the $k$-words that satisfy~\eqref{property} have a regular $m$-block parsing with at least a $(1-\delta/2)(k/m)^d$ words in $C^{\mu}_{m}$. This is by the following lemma. 

\begin{lemma}[Lemma~3.2 in~\cite{UniversallyTypical}]
Let $k\geq 4d m/\delta$. 
If $C\subseteq\Sigma^m$ and $w\in\Sigma^k$ satisfy 
$\sum_{\br\in\Lambda_{k-m+1}}\mathds{1}_{[C]}(\sigma_{\br}[w])\geq (1-\delta/4)(k-m+1)^d$, then, for some $\bp\in\Lambda_m$, 
$\sum_{\substack{ \br \in m\cdot \mathbb{Z}^{d}:  \\ (\Lambda _{m}+\br+\bp)\subseteq \Lambda _{k}}}\mathds{1}_{[C]}(\sigma_{\br + \bp} [w])\geq (1-\delta) (k/m)^d$. 
\end{lemma}

Hence each word satisfying~\eqref{property} is in the typical sampling set $\mathcal{T}^{\mu}_{k}(\delta/2,m)$ defined in~\eqref{eq:longeq1}. 
As discussed above, this applies to at least a $(1-{\delta'})$-fraction of the words $w_1,\ldots, w_I$. 
By Theorem~\ref{thm:sampling0}, the log-cardinality of $\mathcal{T}^{\mu}_{k}(\delta/2,m)$ is upper bounded by $k^d(h+\delta)$. 
\end{IEEEproof}

% % LONG EQUATION
%\newcounter{MYtempeqncnt}
\begin{figure*}[!t]
\normalsize
\setcounter{MYtempeqncnt}{\value{equation}}
\setcounter{equation}{7}
\begin{equation}
|C(\tilde L_{k'}^{k^\ast})| 
\leq 
  (1-\epsilon)\tilde{\tilde M}_{k'}^{k^\ast} \Big( \underbrace{\log (\sum_{k=1}^{k^\ast} \tilde M_k)}_{1)} + \underbrace{\log (k^\ast)}_{2)}  + \underbrace{(k^\ast -k')^{d-1}d}_{3)}  \Big)
   + \epsilon \tilde{\tilde M}_{k'}^{k^\ast} \Big( \underbrace{\log (\sum_{k=1}^{k^\ast} \tilde M_k)}_{1)} + \underbrace{\log (k^\ast)}_{2)}  + \underbrace{{k^\ast}^d}_{3)}  \Big). \label{eq:longeq}
\end{equation}
\setcounter{equation}{\value{MYtempeqncnt}}
\hrulefill
\vspace*{4pt}
\end{figure*}

The next lemma addresses the ``nestedness'' of the parsings generated by Algorithm~\ref{alg3}, that is, the amount of sub-words of parsed words, which are sub-words of previously parsed words. 
Consider the sequence $(k^\ast_i(\alpha_0))$ of all $k$ for which $L_k$ has the typical number of distinct parsed words, $A^{k^d(h-\alpha_0)} \leq M_k \leq A^{k^d(h+\alpha_0)}$, for some $\alpha_0>0$. 

\begin{lemma}
\label{lemma:shortpointer}
For all $\epsilon,\alpha',\alpha$ with $1/2>\alpha'>\alpha>0$, 
there is an $\alpha_0<\alpha$ and an $i_0\in\mathbb{N}$ such that for all $i\geq i_0$, 
at least a $(1-\epsilon)$-fraction of all words parsed in $L_{k'}^{k^\ast_i(\alpha_0)}$, $k'=\lfloor(1-\alpha)k_i^\ast(\alpha_0) \rfloor$, have a leading sub-word of side-length larger or equal to $k''=\lfloor(1-\alpha')k_i^\ast(\alpha_0)\rfloor$ that is a leading sub-word of a previously parsed word. 
\end{lemma}

\begin{figure}
\centering
\includegraphics[clip=true,trim=0.1cm .6cm .5cm .3cm, scale=1]{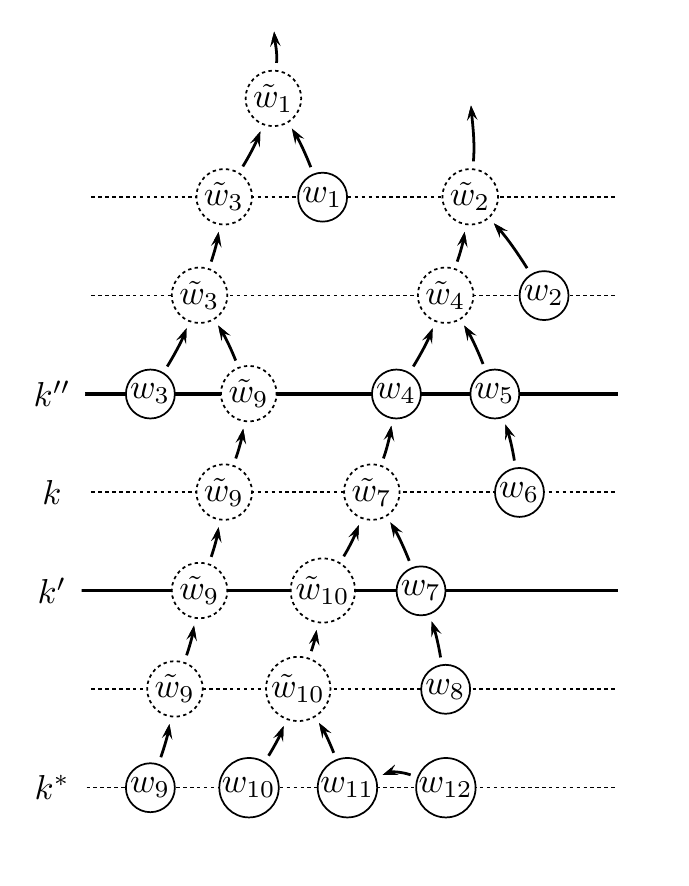}
\caption{Sketch of the tree structure of nested words parsed by Algorithm~\ref{alg3}. 
Most words parsed in $L_{k'}^{k^\ast}$ have leading sub-words that are leading sub-words of words parsed in $L_{k''}^{k^\ast}$. 
Assuming the contrary would imply the existence of too many distinct leading $k''$-sub-words. 
In the figure, the horizontal lines stand for the side-lengths of the parsed words, $w_i$ represents the $i$-th parsed word, 
and a $\tilde w_i$ in the horizontal line $k$ represents the leading $k$-sub-word of $w_i$ in the case that it is not a leading $k$-sub-word of any previously parsed word $w_j$, $j<i$. 
The arrows represent the pointers defined in~\ref{section:coding-algorithm} item~1. 
For example, the arrow from $\tilde w_{10}$ to $\tilde w_7$ signifies that $p_{10}=7$ and that the considered leading sub-word has side-length~$k$.}
\label{fig:tree}
\end{figure}

\begin{IEEEproof} %[Proof of Lemma~\ref{lemma:shortpointer}]
Assuming that an $\epsilon$-fraction of all words parsed in $L_{k'}^{k^\ast}$ do not have leading sub-words of side-length larger or equal to $k''$ that are leading sub-words of previously parsed words, 
we show that there are too many distinct non-overlapping sub-words covering a positive fraction of~$\Lambda_{n(k^\ast)}$, contradicting Theorem~\ref{thm:sampling}. 

We abbreviate $k_i^\ast(\alpha_0)$ by $k^\ast$. 
The set of leading $k''$-sub-blocks of the blocks parsed in $L_{k'}^{k^\ast}$ has a total volume at least $(1-\alpha') |L_{k'}^{k^\ast}|$. 
According to Proposition~\ref{proposition:lastLs}, $|L_1^{k'-1}| \leq \tilde\epsilon |L_{k'}^{k^\ast}|$, 
where $\tilde\epsilon$ becomes arbitrarily small when $k^\ast$ is large enough.  
Hence $(1-\alpha') |L_{k'}^{k^\ast}| \geq ((1-\alpha') - \tilde\epsilon) |L_1^{k^\ast}|$.  
In fact, the parsing contains a set of non-overlapping leading $k''$-sub-blocks that covers at least a $((1-\alpha') - \tilde\epsilon)$-fraction of the region~$\Lambda_{n(k^\ast)}$.

By Lemma~\ref{proposition:typicalsizeL}, the number $\tilde M_{k'}^{k^\ast}$ of words parsed in $L_{k'}^{k^\ast}$ is lower bounded by $A^{k'^d(h-\alpha_0)}$. 
Assume that an $\epsilon$-fraction of these words do not have a leading $k''$-sub-word that is a leading sub-word of any previously parsed word. 
Each of these words must have a distinct leading $k''$-sub-word, since otherwise some of them would be sub-words of previously parsed words. 
The situation is illustrated in Figure~\ref{fig:tree}. 
Under this assumption, the number of distinct leading $k''$-sub-words is at least $\epsilon \tilde M_{k'}^{k^\ast}$. 
Keep in mind that $\tilde M_{k'}^{k^\ast}\geq A^{k'^d(h-\alpha_0)}$. 

On the other hand, choosing $k^\ast$ large enough, $\tilde\epsilon$ becomes arbitrarily small, and $((1-\alpha') - \tilde\epsilon)>0$. 
Recall that the latter lower bounds the fraction of $\Lambda_{n(k^\ast)}$ covered by leading $k''$-sub-blocks in $L_{k'}^{k^\ast}$. 
By Theorem~\ref{thm:sampling}, using $\beta=\epsilon((1-\alpha')-\tilde\epsilon)$, there is a set $\mathcal{T}_{k''}$ of log-cardinality $\log|\mathcal{T}_{k''}| \leq {k''}^d(h+\delta)$, 
which contains at least a $(1-\delta)$-fraction of all leading $k''$-sub-words in~$L_{k'}^{k^\ast}$.

Hence $A^{k''^d(h+\delta)} \geq  (\epsilon-\delta) A^{k'^d(h-\alpha_0)}$. 
This is a contradiction when $\delta<\epsilon$ and 
$\frac{(1-\alpha')^d}{(1-\alpha)^d}\frac{(h+\delta)}{(h-\alpha_0)}<1$, which is always the case for sufficiently small $\delta$ and $\alpha_0$. 
\end{IEEEproof}

We have now developed all the tools that we need to complete the proof of the main theorem. 

\begin{IEEEproof}[Proof of Theorem~\ref{theorem:optimality}] 
Consider a word $w$ parsed at a given time step. 
Let $\tilde w$ be the largest leading sub-word of $w$ that is the leading sub-word of a previously parsed word. 
The code of $w$ consists of the following: 

\begin{enumerate}
\item A pointer to the first previous word $w'$ which contains $\tilde w$ as a leading sub-word. %See Figure~\ref{fig:pointers}. 
\item A descriptor of the side-length of $\tilde w$. 
\item An encoding of the boundary entries of $w$, i.e., the entries of $w$ that are not contained in~$\tilde w$.  
\end{enumerate}

Consider the sequence $K=(k_i^\ast)$ of all $k$ for which $L_k$ has the typical number of blocks, according to Lemma~\ref{proposition:typicalsizeL}, for some $\alpha_0>0$. 
Let us first assume that at the current time step, 
the parsing block side-length is equal to some $k^\ast\in K$. 
Let $n^\ast\in\mathbb{N}$ denote the side-length of the smallest cubical region containing the $L$ region being parsed at the current time step. 
We have $n(k^\ast-1)\leq n^\ast \leq n(k^\ast)$. 

Let $\tilde L_{k^\ast}$ denote the set $L_{k^\ast}\cap\Lambda_{n^\ast}$. 
Let $\tilde{\tilde M}_{k^\ast}$ denote the total number of $k^\ast$-blocks in $\tilde L_{k^\ast}$. 
The region $\tilde L_{k^\ast}$ is contained in $L_{k^\ast}$ and hence $\tilde{\tilde M}_{k^\ast}\leq \tilde M_{k^\ast}$.
The number of blocks of side-length from $k'$ to $k^\ast$ parsed in $\Lambda_{n^\ast}$ is given by 
$\tilde{\tilde M}_{k'}^{k^\ast}:=\tilde M_{k'}^{(k^\ast-1)} + \tilde{\tilde M}_{k^\ast}$. 
By Lemma~\ref{lemma:shortpointer}, a $(1-\epsilon)$-fraction of the words parsed in $\tilde L_{k'}^{k^\ast}$ have large nested sub-words. 
For the encoding $C(\tilde L_{k'}^{k^\ast})$ of the words parsed in $\tilde L_{k'}^{k^\ast}$, we have the bound~\eqref{eq:longeq}. \addtocounter{equation}{1}
By similar arguments as in~\eqref{eq:upperbd}, 
$\tilde{\tilde M}_{k'}^{k^\ast}\leq \sum_{k=1}^{k^\ast} \tilde M_k \leq c A^{ {k^\ast}^d(h+\alpha_0)}$, 
and hence  
\ifpreprint
\begin{align}
|C(\tilde L_{k'}^{k^\ast})|  \leq 
& (1-\epsilon)\tilde{\tilde M}_{k'}^{k^\ast} \left({k^\ast}^d(h+\alpha_0)+ o({k^\ast}^d) \right)  \nonumber \\
& + \epsilon \tilde{\tilde M}_{k'}^{k^\ast} (h+1){k^\ast}^d \nonumber\\
\leq &(1+\epsilon') {n^\ast}^d (h+\alpha_0) + \epsilon {n^\ast}^d . \label{eq:rhso}
\end{align}
\else
\begin{align}
&|C(\tilde L_{k'}^{k^\ast})| \nonumber\\
& \;\leq (1-\epsilon)\tilde{\tilde M}_{k'}^{k^\ast} \left({k^\ast}^d(h+\alpha_0) + o({k^\ast}^d)  \right) + \epsilon \tilde{\tilde M}_{k'}^{k^\ast} (h+1){k^\ast}^d \nonumber\\
& \;\leq (1+\epsilon') {n^\ast}^d (h+\alpha_0) + \epsilon {n^\ast}^d . \label{eq:rhso}
\end{align}
\fi
Since $\epsilon'$ and $\alpha_0$ can be chosen arbitrarily small, the right hand side~\eqref{eq:rhso} can be given as ${n^\ast}^d h (1+ o(1))$. 
Since $|L_1^{k'-1}|$ is arbitrarily small compared with $|L_{k'}^{k^\ast}|$ as $k^\ast$ tends to infinity, we obtain 
\begin{equation*}
|C(x^{n^\ast})|  \leq  {n^\ast}^d h (1+ o(1)).   
\end{equation*}
Since the relative volume covered by regions $L_k$ with $k\not\in K=(k_i^\ast)$ tends to zero (see~Lemma~\ref{proposition:typicalsizeL}), the code length of these regions is negligible. 
Hence, for arbitrary $n$ we get the same bound as for $n^\ast$, 
\begin{equation*}
|C(x^{n})|  \leq  n^d h (1+ o(1)).   
\end{equation*}
This completes the proof. 
\end{IEEEproof}

\section{Remarks}
\label{section:remarks}
Our proof of the asymptotic optimality of Algorithm~\ref{alg3} also yields the optimality of the modified algorithm that considers arbitrary cubical sub-words of parsed words instead of leading cubical sub-words alone. 
This additional freedom in the choice of the pointers can only improve the compression ratio. 

Our proof shows the asymptotic optimality for any choice of the parameter $\phi$. 
Nevertheless, this parameter may well influence the velocity of convergence of the compression ratio. 
Fine tuning of our algorithm may be of interest for practical implementations. 
An extensive empirical study is in preparation. 

The $1$-dimensional special case of our algorithm differs from the classical LZ algorithm, since our algorithm allows word repetitions and uses a strictly non-decreasing parsing block size. 
The LZ algorithm can be obtained from our algorithm by setting $\phi=0$, 
defining the boundary regions $L$ as $\{m+1,\ldots, k\}$, instead of $\Lambda_{\lfloor m/k \rfloor k+k}\setminus \Lambda_{\lfloor m/k \rfloor k}$, and evaluating the recurrence quantifying function $J$ not on the current parsing but ``predictively'' on the parsing that one would obtain in the next iteration for a given choice of the parsing block size $k$ for the current iteration. 
Our proofs do not directly apply to that setting. 
However, modifications evaluating $J$ predictively and allowing shrinking parsing block size are plausible.

\section*{Acknowledgment}

We would like to thank the Information Theory of Cognitive Systems Group at the Max Planck Institute MIS for funding and infrastructure for this project.

\bibliographystyle{IEEEtran}
\bibliography{references}

\end{document}